\documentclass[12pt]{article} 
\usepackage{graphicx}
\def \d {{\rm d}}

\def \bt {{\bar{t}}}
\def \brho {{\bar{\rho}}}
\def \bz {{\bar{z}}}
\def \bphi {{\bar{\phi}}}

\def\boxit#1{\vbox{\hrule\hbox{\vrule\kern3pt
              \vbox{\kern3pt#1\kern3pt}\kern3pt\vrule}\hrule}}

\begin{document}

\title{\bf Global aspects of accelerating and rotating black hole space-times}

\author{J. B. Griffiths$^1$\thanks{E--mail: {\tt J.B.Griffiths@Lboro.ac.uk}} \ 
and J. Podolsk\'y$^2$\thanks{E--mail: {\tt Podolsky@mbox.troja.mff.cuni.cz}}
\\ \\ \small
$^1$Department of Mathematical Sciences, Loughborough University, \\
\small Loughborough,  Leics. LE11 3TU, U.K. \\
\small $^2$Institute of Theoretical Physics, Charles University in Prague,\\
\small V Hole\v{s}ovi\v{c}k\'ach 2, 18000 Prague 8, Czech Republic.}

\date{\today}
\maketitle

\begin{abstract}
\noindent
The complete family of exact solutions representing accelerating and rotating black holes with possible electromagnetic charges and a NUT parameter is known in terms of a modified Pleba\'nski--Demia\'nski metric. This demonstrates the singularity and horizon structure of the sources but not that the complete space-time describes two causally separated black holes. To demonstrate this property, the metric is first cast in the Weyl--Lewis--Papapetrou form. After extending this up to the acceleration horizon, it is then transformed to the boost-rotation-symmetric form in which the global properties of the solution are manifest. The physical interpretation of these solutions is thus clarified. 
\end{abstract}


\section{Introduction}

It is known that the Pleba\'nski--Demia\'nski \cite{PleDem76} family of solutions includes the case for an accelerating and rotating black hole. This family is characterised by a related pair of quartic functions which, for this case to occur, must possess four distinct roots. 
When considering solutions of this type with a zero cosmological constant, it had become traditional to use a coordinate freedom to remove the linear terms in these quartic functions. 
This was thought to remove the NUT parameter.
However, Hong and Teo \cite{HongTeo03} have shown that the available freedom is much better used to simplify the roots of these functions. Then, in the recent extension of their analysis \cite{HongTeo05} to the rotating case, they obtained a new solution for an accelerating and rotating black hole which differs from what is usually called the ``spinning $C$-metric'' \cite{FarZim80b, BicPra99, Pravdas02} in which the linear terms are set to zero. 
Surprisingly, it is the new solution of Hong and Teo which represents the NUT-free case, while the older ``spinning $C$-metric'' actually retains NUT-like properties (i.e. part of the axis corresponds to a ``torsion'' singularity which is surrounded by a region that contains closed timelike lines). 
A different and more general form of the metric was presented in \cite{GriPod05} which confirms that the ``spinning $C$-metric'' does indeed possess an effective non-zero NUT parameter.

The form of the metric presented in \cite{GriPod05} covers the complete family of exact solutions which represent accelerating and rotating black holes with possible electromagnetic charges and an arbitrary NUT parameter. 
This form of the metric clearly describes the internal horizon and singularity structure of a single black hole as far as its associated acceleration horizon. However, for accelerating black holes with no rotation, no charges and no NUT parameter, the metric reduces to that of the familiar $C$-metric. 
In this particular case, it is known that the complete analytic extension of the space-time contains two causally separated black holes which accelerate away from each other under the action of forces which are represented by conical singularities along appropriate parts of the axis of symmetry. 
For the general family of solutions described in \cite{GriPod05}, topological singularities of the required type occur on the axis of symmetry. However, the existence of a second black hole cannot be demonstrated explicitly using this form of metric in which the coordinates do not extend significantly beyond the acceleration horizon.

The purpose of the present paper is to continue to explore the physical interpretation of this family of solutions, particularly by finding an analytic extension of the space-time beyond the acceleration horizon. 
This will be achieved for the general case which, in addition to a rotation, also generally includes electromagnetic charges and an arbitrary NUT parameter. 
As a first step, it is necessary to express the metric for the stationary region outside the black hole horizon in Weyl--Lewis--Papapetrou form. However, the metric that is obtained in a simple way does not even cover this region completely. The next step is therefore to re-cast this form of the metric in the unique form which can be extended throughout the entire stationary region between the outer black hole event horizon and the acceleration horizon. Finally, the metric is transformed to boost-rotation-symmetric form which covers both stationary and time-dependent regions. This provides the maximum analytic extension of the space-time outside the black hole horizons, and confirms that these solutions indeed generally describe a pair of accelerating black holes (with other attributes) as expected.

For the case of the spinning $C$-metric, which actually includes a specific non-zero value for the NUT parameter but no charges, the procedure for expressing the metric in boost-rotation-symmetric coordinates has been described in detail by Bi\v{c}\'ak and Pravda \cite{BicPra99}. 
However, it was conjectured by Hong and Teo \cite{HongTeo05} that this procedure would be simplified using the newer forms of the metric in which the expressions for the roots are much more convenient. It is demonstrated here that this is indeed the case. Including also charges and a general NUT parameter, it is shown in detail how the general solution in boost-rotation-symmetric coordinates can be derived. 
Particular attention is paid to the case in which the NUT parameter is zero, as this represents the physically more significant case for an accelerating, rotating and charged black hole. Using this approach, it is also shown that, when the rotation and charges vanish, the resulting expressions for the $C$-metric in boost-rotation-symmetric coordinates reduce to remarkably simple forms.

\section{Modified forms of the Pleba\'nski--Demia\'nski metric} 

An appropriate starting point is the Pleba\'nski--Demia\'nski metric which describes aligned type~D electrovacuum space-times in which the repeated principal null congruences are expanding and for which the orbits of the Killing vectors are non-null. In a previous paper \cite{GriPod05}, we found it convenient to start with a modified form of this metric. For the case in which the cosmological constant is set to zero, this is given by 
  \begin{equation}
  \begin{array}{r}
{\displaystyle  \d s^2={1\over(1-\alpha pr)^2} \Bigg[
{Q\over r^2+\omega^2p^2}(\d\tau-\omega p^2\d\sigma)^2
  -{P\over r^2+\omega^2p^2}(\omega\d\tau+r^2\d\sigma)^2 } \hskip3pc \\[12pt]
  {\displaystyle -{r^2+\omega^2p^2\over P}\,\d p^2
-{r^2+\omega^2p^2\over Q}\,\d r^2 \Bigg],}
  \end{array}
  \label{PleDemMetric}
  \end{equation}
  where
  \begin{equation}
  \begin{array}{l}
  P =k +2\omega^{-1}np -\epsilon p^2 +2\alpha mp^3-\alpha^2(\omega^2 k+e^2+g^2)p^4, \\[8pt]
  Q =(\omega^2k+e^2+g^2) -2mr +\epsilon r^2 -2\alpha\omega^{-1}nr^3-\alpha^2kr^4.
  \end{array}
  \label{PQeqns}
  \end{equation}
 and $m$, $n$, $e$, $g$, $\alpha$, $\omega$, $\epsilon$ and $k$ are constants. The parameter $\alpha$ represents the acceleration of the sources, while $\omega$ is proportional to the twist of the repeated principal null directions and this relates to both the Kerr-like angular velocity of the sources and the NUT-like properties of the space-time. 
For certain choices of the other parameters, $m$ is related to the mass of the source and $n$ to the NUT parameter. The parameters $e$ and $g$ denote the electric and magnetic charges of the sources. 
Finally, $\epsilon$ and $k$ are parameters which are related to the curvature of certain 2-surfaces and to the choice of coordinates on them. They can each be scaled to any value (without changing their signs) using the remaining freedoms. 
To retain Lorentzian signature, it is necessary that $P>0$, so the coordinate $p$ is assumed to range between appropriate roots of the quartic~$P(p)$. For black hole solutions, $P$ must have at least two roots which can be specified for convenience by an explicit choice of the parameters $\epsilon$ and $k$. Horizons occur at the roots of the quartic~$Q(r)$.

To introduce an explicit NUT parameter into the metric (\ref{PleDemMetric}), it is necessary to include a shift in the coordinate~$p$. Specifically, as explained in \cite{GriPod05}, we perform the coordinate transformation 
  \begin{equation}
  p=\omega^{-1}(l+a\cos\theta), \qquad  \tau=\bar t-(l+a)^2a^{-1}\bphi, 
  \qquad \sigma=-\omega a^{-1}\bphi,
  \label{trans1A}
  \end{equation}
  where $a$ and $l$ are arbitrary parameters, which are chosen such that the roots of $p$ occur at $\theta=0$ and $\theta=\pi$. In this case, it is appropriate to express the parameters $n$ and $\epsilon$ in terms of $a$ and $l$ as 
 \begin{equation}
 n= {\omega^2k\,l\over a^2-l^2} -{\alpha(a^2-l^2)\over\omega}\,m 
 +{\alpha^2(a^2-l^2)l\over\omega^2}\,(\omega^2k+e^2+g^2) 
 \label{n}  
 \end{equation} 
 and 
 \begin{equation} 
 \epsilon= {\omega^2k\over a^2-l^2}+4{\alpha l\over\omega}\,m 
 -{\alpha^2(a^2+3l^2)\over\omega^2}(\omega^2k+e^2+g^2), 
 \label{epsilon}
 \end{equation}
 and to choose $k$ to satisfy the equation 
  \begin{equation}
  \left( {\omega^2\over a^2-l^2}+3\alpha^2l^2 \right)\,k =1 +2{\alpha l\over\omega}\,m 
  -3{\alpha^2l^2\over\omega^2}(e^2+g^2). 
  \label{k}
  \end{equation} 

 \goodbreak\noindent
 In this way, we obtain the metric  
  \begin{equation}
  \begin{array}{l}
{\displaystyle \d s^2={1\over\Omega^2}\left\{
{Q\over\varrho^2}\left[\d\bar t- \left(a\sin^2\theta
+4l\sin^2{\textstyle{\theta\over2}} \right)\d\bar\phi \right]^2
   -{\varrho^2\over Q}\,\d r^2 \right.
} \\[8pt]
  \hskip8pc {\displaystyle 
 \left. -{\tilde P\over\varrho^2} \Big[ a\d\bar t  
  -\Big(r^2+(a+l)^2\Big)\d\bar\phi \Big]^2  
-{\varrho^2\over\tilde P}\sin^2\theta\,\d\theta^2 \right\}, }
\end{array}
  \label{newMetric}
  \end{equation}
  where 
  \begin{equation}
  \begin{array}{l}
  {\displaystyle \Omega=1-{\alpha\over\omega}(l+a\cos\theta)\,r }\,, \\[6pt]
  \varrho^2 =r^2+(l+a\cos\theta)^2 \,, \\[6pt]
  \tilde P= \sin^2\theta\,(1-a_3\cos\theta-a_4\cos^2\theta) \,, \\[6pt]
  Q= {\displaystyle \left[(\omega^2k+e^2+g^2)\bigg(1+2\alpha {l\over\omega}\,r\bigg) 
  -2mr +{\omega^2k\over a^2-l^2}\,r^2\right] } \\[6pt]
   \hskip5pc \times {\displaystyle 
   \left[1+\alpha{(a-l)\over\omega}\,r\right] \left[1-\alpha{(a+l)\over\omega}\,r\right] } \,,
  \end{array} 
  \label{newMetricFns}
  \end{equation} 
 and 
 \begin{equation}
 \begin{array}{l}
  {\displaystyle a_3= 2\alpha{a\over\omega}m -4\alpha^2{al\over\omega^2}
  (\omega^2k+e^2+g^2) } \,, \\[6pt]
  {\displaystyle a_4= -\alpha^2{a^2\over\omega^2}(\omega^2k+e^2+g^2) } \,,
  \end{array}
 \label{a34}
 \end{equation} 
 with $k$ given by (\ref{k}).

The metric (\ref{newMetric}) is regular at the pole $\theta=0$, which corresponds to an axis, so that $\bar\phi$ can be taken as a periodic coordinate. However, a specific kind of singularity occurs at the other pole $\theta=\pi$. This is consistent with the interpretation that $a$ corresponds to a Kerr-like rotation parameter for which the corresponding metric components are regular on the entire axis, while $l$ corresponds to a NUT parameter for which the corresponding components are only regular on the half-axis $\theta=0$.

The metric in the form (\ref{newMetric}) contains seven arbitrary parameters $m$, $a$, $l$, $e$, $g$, $\alpha$ and $\omega$. Of these, the first six can be varied independently, and the remaining freedom can be used to set $\omega$ to any convenient value if at least one of the parameters $a$ or $l$ are non-zero (otherwise $\omega=0$ automatically).

The metric (\ref{newMetric}) represents the complete family of solutions which describe accelerating and rotating charged black holes with a generally non-zero NUT parameter. It reduces explicitly to known forms of either the Kerr--Newman--NUT class of solutions or the \hbox{$C$-metric} in appropriate cases. If $|l|\le|a|$, there is a curvature singularity when $\varrho^2=0$; i.e. at $r=0$, $\cos\theta=-l/a$. However, if $|l|>|a|$, the metric is non-singular. These properties are typical of the Kerr--NUT black hole.

If $\alpha\ne0$, this solution represents a black hole which accelerates along the axis of symmetry in the direction $\theta=0$. 
However, although it is far from obvious, the complete analytically extended space-time represents a pair of causally separated black holes which are accelerating away from each other in opposite directions. 
To demonstrate this property, which is well known for the special case of the $C$-metric, it is necessary to find an analytic extension of the space-time through the acceleration horizon. For this, we need to re-express the solution in terms of boost-rotation-symmetric coordinates. However, to transform the metric to such a form, it is first convenient to rewrite it in terms of the Weyl--Lewis--Papapetrou metric.

\section{The Weyl--Lewis--Papapetrou form}

Our aim in this section is to express the stationary region of the space-time represented by the metric (\ref{newMetric}) in terms of the Weyl--Lewis--Papapetrou line element
  \begin{equation}
\d s^2= e^{2\bar U}(\d\bt+\bar A\,\d\bphi)^2
 -e^{-2\bar U}\Big[ e^{2\bar\nu}(\d\brho^2+\d\bz^2)+\brho^2\d\bphi^2 \Big] \ ,
  \label{WeylMetric}
  \end{equation}
where $\bar U$, $\bar A$ and $\bar\nu$ are specific functions of the coordinates $\brho$ and $\bz$ only. By using 
 $$ p={l\over\omega}+{a\over\omega}\,\cos\theta $$ 
 as in (\ref{trans1A}), and comparing coefficients in the metrics (\ref{WeylMetric}) and (\ref{newMetric}) for the same coordinates $\bar t$ and $\bar\phi$, it can be seen that 
  \begin{equation}
 e^{2\bar U}={Q-a^2\tilde P\over\varrho^2\,\Omega^2}, \qquad
 \bar A=-{(a+l)^2\over a} +
 {\omega^2p^2Q+a^2r^2\tilde P\over a(Q-a^2\tilde P)}, 
  \label{Weyltrans}
  \end{equation} 
 and \ $\bar\rho^2=\tilde P\,Q\,\Omega^{-4}$, \ in which the right-hand-sides are still expressed at this point in terms of the coordinates $r$ and $p$.

It may be observed that the metric (\ref{WeylMetric}) with (\ref{Weyltrans}) only covers that part of  the stationary region for which $Q>a^2P$. For the non-accelerating but rotating case (i.e. when $\alpha=0$ and $a\ne0$) this is already familiar as the region outside the ergosphere of the Kerr--Newman space-time (i.e. outside the surface of infinite redshift). Thus, when $a\ne0$, the forms (\ref{Weyltrans}) do not apply either through the ergosphere or near the acceleration horizon. This point will be returned to in the following section.

In this restricted part of the stationary region, $\tilde P$ and $Q$ are both positive, and it can be shown that the coordinate transformation between the two forms of the metric (\ref{newMetric}) and (\ref{WeylMetric}) must have the structure
\begin{equation}
  \brho= \frac{\sqrt{\tilde PQ}}{\>\Omega^2}\ , \qquad
  \bz  = {\omega\over a}\frac{\Sigma}{\>\Omega^2} +z_0\ , 
 \label{transf1}  
 \end{equation}
 where 
\begin{eqnarray}
 \Omega(r,p) &=& 1-\alpha pr\ , \nonumber \\
 \Sigma(r,p) &=& \alpha(\omega^2k+e^2+g^2)p^2 -\alpha k r^2 
 +\epsilon pr  - (mp+\omega^{-1}n r) (1+\alpha pr)\ , \nonumber 
 \label{coef1}  
 \end{eqnarray}
 and it is convenient to set 
\begin{equation}
 z_0 = {l\over a}\Big[\,m-\alpha{l\over\omega}(\omega^2k+e^2+g^2)\Big] \,. \label{z0} 
 \end{equation}
 The function $\bar\nu$ is then given by 
  \begin{equation} 
 e^{2\bar U}e^{-2\bar\nu} ={\Omega^2\over\varrho^2}\left( 
 \tilde P\,\bz_{,p}^{\ 2}+Q\,\bz_{,r}^{\ 2} \right). 
  \label{nu} 
  \end{equation}  
 It may be noted that the familiar electrovacuum field equations for the Weyl--Lewis--Papapetrou metric are automatically satisfied by the quartic structure of the functions $\tilde P(p)$ and $Q(r)$. However, it may be observed that the transformation (\ref{transf1}) does not apply directly when $a=0$ unless in addition $\omega=0$.

In order to explicitly express the metric functions $\bar U$, $\bar A$ and $\bar\nu$ in terms of $\brho$ and $\bz$, it is necessary to invert the relations (\ref{transf1}). To achieve this, it is convenient as in \cite{Bonnor83,BicPra99,HongTeo03} to introduce the functions
 \begin{equation}
 R_i=\sqrt{\brho^2+(\bz-\bz_i)^2}\ ,\qquad i=1,2,3\ , 
 \label{Ri}  
 \end{equation}
 where $\bz_i$ are constants which are specifically chosen in such a way that further constants $A_i$, $B_i$ and $C_i$ can be found such that 
  \begin{equation}
 A_i\,p +B_i\,r +C_i(1+\alpha pr) =(1-\alpha pr)R_i \,.
 \label{RiE}  
 \end{equation}
 From the equations (\ref{transf1}), (\ref{Ri}) and (\ref{RiE}), we obtain explicit conditions for the constant parameters, namely
 \begin{eqnarray}
  A_i^2 &=& {\omega^2\over a^2}m^2 -{\omega^2\over a^2}(\omega^2k+e^2+g^2) \Big(\epsilon+2\alpha {a\over\omega} (\bz_i-z_0)\Big)\ , \nonumber \\
  B_i^2 &=& {n^2\over a^2} + {\omega^2\over a^2}k\Big(\epsilon+2\alpha {a\over\omega} (\bz_i-z_0)\Big)\ , \label{ci2} \\
  C_i^2 &=& {\omega^2\over a^2}(\omega^2k+e^2+g^2) k +(\bz_i-z_0)^2\ , \nonumber 
 \end{eqnarray}
and
 \begin{eqnarray}
  A_iB_i &=& -{\omega\over a^2}n m 
  -\Big({\omega\over a}\epsilon+2\alpha(\bz_i-z_0)\Big)(\bz_i-z_0)\ , \nonumber \\
  A_iC_i &=& {\omega\over a^2}(\omega^2k+e^2+g^2)n +{\omega\over a}m(\bz_i-z_0)\ , \label{cicj} \\
  B_iC_i &=& -{\omega^2\over a^2}km+ {n\over a} (\bz_i-z_0). \nonumber 
 \end{eqnarray}
 It can be easily shown that the system of conditions (\ref{ci2})--(\ref{cicj}) is only consistent provided $Z_i=a\omega^{-1}(\bz_i-z_0)$ is the solution of the cubic equation 
 $$ \begin{array}{l}
 2\alpha Z_i^3+\epsilon Z_i^2+2[\omega^{-1}n m+\alpha k(\omega^2k+e^2+g^2)]Z_i 
 \\[6pt]
 \hskip20mm +[(\epsilon k+\omega^{-2}n^2)(\omega^2k+e^2+g^2)-km^2]=0\ .
\end{array} $$ 
 If this condition is satisfied, the constants $A_i$, $B_i$ and $C_i$ are uniquely defined by (\ref{ci2})--(\ref{cicj}) up to a choice of signs, which is then fixed by the condition that $R_i>0$ in (\ref{RiE}) for appropriate ranges of $p$ and $r$.

 Remarkably, when the expressions for $n$, $\epsilon$, $k$ and $z_0$ from (\ref{n}), (\ref{epsilon}), (\ref{k}) and (\ref{z0}) are inserted into this cubic, it can be seen to factorize most conveniently in the form 
\begin{equation}
(2\alpha \bz_i+K)(\bz_i^2-L^2)=0 \ ,
\label{cubic}
\end{equation}
 where 
 $$ \begin{array}{l}
  K = {\displaystyle \frac{\omega}{a}-\alpha^2\frac{a}{\omega}(\omega^2k+e^2+g^2) }\ , \\[10pt]
  L^2 = {\displaystyle \Big[\, m-2\alpha\frac{l}{\omega}(\omega^2k+e^2+g^2)\,\Big]^2
  -(\omega^2k+e^2+g^2) }\ .
 \end{array} $$ 
This explicitly identifies the roots, which we take in the order  
 \begin{equation} 
 \begin{array}{l} 
  \bz_1 = {\displaystyle + \sqrt{
  \Big[\, m-2\alpha\frac{l}{\omega}(\omega^2k+e^2+g^2)\,\Big]^2
    -(\omega^2k+e^2+g^2)}  }\ ,  \\[10pt]
  \bz_2 = {\displaystyle - \sqrt{
  \Big[\, m-2\alpha\frac{l}{\omega}(\omega^2k+e^2+g^2)\,\Big]^2
    -(\omega^2k+e^2+g^2)} }\ ,  \\[10pt]  
  \bz_3 = {\displaystyle -\frac{\omega}{2\alpha a} 
  +\frac{\alpha a}{2\omega}(\omega^2k+e^2+g^2) }\ . 
 \end{array}
 \label{Zigen}  
 \end{equation} 
 With these roots, and when $\alpha\ne0$, the definitions (\ref{Ri}) lead to the identity 
 \begin{equation}
 (\bz_1+\bz_3)R_1^2 +(\bz_1-\bz_3)R_2^2 -2\bz_1R_3^2 
=2\bz_1(\bz_1^2-\bz_3^2) \,, 
 \label{Risquared}
 \end{equation} 
 which will be particularly useful later.

To invert the transformation (\ref{transf1}), two independent equations are required. Thus, we only need two independent components of the equation (\ref{RiE}), and thus only two distinct roots of (\ref{cubic}) are necessary. However, in general, the equation (\ref{cubic}) has the above three distinct roots, and these give up to three distinct components for each of the sets of constants $R_i$, $A_i$, $B_i$ and $C_i$ through (\ref{Ri}), (\ref{ci2}) and (\ref{cicj}). Using the familiar notation for three-dimensional cartesian vectors, equation (\ref{RiE}) can be rewritten in the form 
 \begin{eqnarray}
 {\bf A}\,p +{\bf B}\,r 
 +\left({\bf R}+{\bf C}\right)\alpha pr 
 ={\bf R}-{\bf C} \,.
 \label{vectorR}
 \end{eqnarray} 
 For the case in which (\ref{cubic}) has three distinct roots, the term $\alpha pr$ can initially be treated as an independent variable, so that the three components of (\ref{vectorR}) can be regarded as independent linear equations and their solution can be expressed as 
  \begin{equation} 
 p ={2\>({\bf B}\times{\bf C}\cdot{\bf R}) \over 
 ({\bf A}\times{\bf B}\cdot{\bf C})+({\bf A}\times{\bf B}\cdot{\bf R})} \,, \qquad
 r=-{2\>({\bf A}\times{\bf C}\cdot{\bf R}) \over 
 ({\bf A}\times{\bf B}\cdot{\bf C})+({\bf A}\times{\bf B}\cdot{\bf R})}, 
  \label{pr} 
  \end{equation}  
 where ${\bf R}={\bf R}(\brho,\bz)$ is given by (\ref{Ri}). It can then be verified that the resulting expression for $\alpha pr$ is consistent with this solution.

With the above roots (\ref{Zigen}) for $\bz_i$, we can determine explicitly the components of ${\bf A}$, ${\bf B}$ and ${\bf C}$ as well as ${\bf R}$. Hence, we evaluate the old coordinates (\ref{pr}) in terms of $\brho$ and $\bz$ and substitute them into the metric functions (\ref{newMetricFns}) with $\cos\theta=a^{-1}(\omega p-l)$. Finally, substituting these expressions into (\ref{Weyltrans}) and (\ref{nu}), the transformation to the Weyl--Lewis--Papapetrou form is complete.

Let us now describe two particularly important cases. First we consider the case in which the acceleration $\alpha$ vanishes. Then we will concentrate on the case for accelerating and rotating black holes for which the NUT parameter $l$ is zero.

\subsection{The non-accelerating case}

In the case in which ${\alpha=0}$ (which implies that ${n=l}$, ${\epsilon=1}$ and ${\omega^2k=a^2-l^2}$, see (\ref{n})--(\ref{k})), the equation (\ref{cubic}) reduces to a quadratic which can immediately be solved, yielding the two roots
 \begin{eqnarray}
  \bz_1 &=& +\sqrt{m^2+l^2-(a^2+e^2+g^2)} \ , \nonumber  \\[3pt]
  \bz_2 &=& -\sqrt{m^2+l^2-(a^2+e^2+g^2)} \ . \nonumber
 \end{eqnarray}
Using (\ref{ci2}), (\ref{cicj}) we obtain the corresponding coefficients
 \begin{eqnarray} 
  A_1 &=& -{\omega\over a}\sqrt{m^2+l^2-(a^2+e^2+g^2)} \ , \nonumber \\[3pt]
  B_1 &=& 1 \ , \nonumber  \\
  C_1 &=& -m +{l\over a}\sqrt{m^2+l^2-(a^2+e^2+g^2)} \ , \nonumber 
 \end{eqnarray}
 and
 \begin{eqnarray}
  A_2 &=& {\omega\over a}\sqrt{m^2+l^2-(a^2+e^2+g^2)} \ , \nonumber \\[3pt]
  B_2 &=& 1 \ , \nonumber  \\
  C_2 &=& -m -{l\over a}\sqrt{m^2+l^2-(a^2+e^2+g^2)} \ . \nonumber 
 \end{eqnarray}
Consequently, the equations (\ref{RiE}) become
 \begin{eqnarray}
  -\sqrt{m^2+l^2-(a^2+e^2+g^2)}\cos\theta +(r-m) &=&  R_1\ , \nonumber \\
  \sqrt{m^2+l^2-(a^2+e^2+g^2)}\cos\theta +(r-m) &=&  R_2 \ , \nonumber
 \end{eqnarray} 
 which have the solution 
 \begin{equation} 
 r=m+{R_1+R_2\over2}, \qquad 
 \cos\theta={R_2-R_1\over2\sqrt{m^2+l^2-(a^2+e^2+g^2)}}. 
 \label{KNNUTcoords} 
 \end{equation} 
For this case, (\ref{Weyltrans}) and (\ref{nu}) become 
 \begin{eqnarray}
 e^{2\bar U} &=& {(r-m)^2+a^2\cos^2\theta+e^2+g^2-m^2-l^2 \over
 r^2+(l+a\cos\theta)^2} \,, \nonumber \\[8pt]
 \bar A &=& (1-\cos\theta){[a(1+\cos\theta)+2l](2mr-e^2-g^2)
 +2l[(a+l)(l+a\cos\theta)-r^2]
 \over (r-m)^2+a^2\cos^2\theta+e^2+g^2-m^2-l^2} , \nonumber \\[8pt]
 e^{2\bar U}e^{-2\bar\nu} &=& {(r-m)^2+(a^2+e^2+g^2-m^2-l^2)\cos^2\theta 
 \over r^2+(l+a\cos\theta)^2} \,. \nonumber 
 \end{eqnarray}
 Substituting (\ref{KNNUTcoords}) into these, gives the Weyl--Lewis--Papapetrou form of the Kerr--Newman--NUT family of solutions.

\subsection{The case with vanishing NUT parameter}
\label{noNUT}

Now consider the other physically significant case in which ${l=0}$. This describes accelerating and rotating black holes without any NUT-like properties. 
The metric for this situation, which differs from the so-called ``spinning $C$-metric'' in which $n=0$, was first found by Hong and Teo \cite{HongTeo05} using different coordinates and clarified in \cite{GriPod05}. 
In this case, the condition (\ref{k}) implies that \ ${\omega^2k=a^2}$ \ and we are free to put $\omega=a$. Then, with the conditions (\ref{n}) and (\ref{epsilon}), we obtain 
 $$ \epsilon=1-\alpha^2(a^2+e^2+g^2), \qquad k=1, \qquad n=-\alpha am, $$ 
 and the cubic equation (\ref{cubic}) has the roots (\ref{Zigen}): 
 \begin{equation}
 \begin{array}{l}
  \bz_1 = {\displaystyle -\bz_2\ =\  \sqrt{m^2-(a^2+e^2+g^2)} }\ ,  \\[6pt]
  \bz_3 = {\displaystyle -\frac{1}{2\alpha}+\frac{\alpha}{2}(a^2+e^2+g^2) }\ . 
 \end{array} 
 \label{ZiNL}
 \end{equation} 
 Notice that, in this case, the roots $\bz_1$ and $\bz_2$ are independent of the acceleration parameter~$\alpha$. (This provides considerable simplifications later.) The corresponding sets of coefficients are 
 \begin{eqnarray}
  A_1 &=& \alpha(a^2+e^2+g^2)-\sqrt{m^2-(a^2+e^2+g^2)} \ , \nonumber \\
  B_1 &=& 1+\alpha \sqrt{m^2-(a^2+e^2+g^2)} \ , \nonumber  \\ 
  C_1 &=& -m \ , \nonumber  
 \end{eqnarray}
 and
 \begin{eqnarray}
  A_2 &=& \alpha(a^2+e^2+g^2)+\sqrt{m^2-(a^2+e^2+g^2)} \ , \nonumber \\
  B_2 &=& 1-\alpha \sqrt{m^2-(a^2+e^2+g^2)} \ , \nonumber \\ 
  C_2 &=& -m \ , \nonumber  
 \end{eqnarray}
 and
 \begin{eqnarray}
  A_3 &=& -m \ , \nonumber \\
  B_3 &=& -\alpha m \ , \nonumber \\ 
  C_3 &=& \frac{1}{2\alpha} +\frac{\alpha}{2}(a^2+e^2+g^2) \ . \nonumber 
 \end{eqnarray}

Evaluating the triple scalar products of the vectors {\bf A}, {\bf B}, {\bf C} and {\bf R} and substituting into equations~(\ref{pr}), the desired coordinate transformation can be expressed in the form 
 \begin{equation}
 \begin{array}{l} 
 \ {\displaystyle p = { (\bz_1+\bz_3)({1\over\alpha}+\bz_1)R_1
 +(\bz_1-\bz_3)({1\over\alpha}-\bz_1)R_2 +2m\bz_1\,R_3 \over
 m(\bz_1+\bz_3)R_1 +m(\bz_1-\bz_3)R_2
 +2\bz_1({1\over\alpha}+\bz_3)R_3 
 -2\bz_1(\bz_1^2-\bz_3^2)} } \,, \\[14pt]
 {\displaystyle \alpha r = 
 { (\bz_1+\bz_3)({1\over\alpha}-\bz_1+2\bz_3)R_1 
 +(\bz_1-\bz_3)({1\over\alpha}+\bz_1+2\bz_3)R_2 +2m\bz_1\,R_3 \over
 m(\bz_1+\bz_3)R_1 +m(\bz_1-\bz_3)R_2
 +2\bz_1({1\over\alpha}+\bz_3)R_3 -2\bz_1(\bz_1^2-\bz_3^2)} } \,. 
 \end{array} 
 \label{przerol}
 \end{equation} 
 in which $\bz_1$ and $\bz_3$ are given in (\ref{ZiNL}). 
 The limit of this as $\alpha\to0$ is consistent with the previous result (\ref{KNNUTcoords}) with $l=0$. It is significant that the meaning of $R_1$ and $R_2$ is the same in both cases, while $R_3$ and $-\bz_3$ here approach infinity as $\alpha\to0$.

The metric functions (\ref{Weyltrans}) and (\ref{nu}) are now given by 
  \begin{equation}
 e^{2\bar U}={Q-a^2\tilde P\over\varrho^2\,\Omega^2}, \qquad
 \bar A=-a\,{(1-p^2)Q -(r^2+a^2)\tilde P\over Q-a^2\tilde P}, \qquad
 e^{2\bar U}e^{-2\bar\nu}={F^2Q+G^2\tilde P\over\varrho^2\,\Omega^4},
 \label{Weyltrans2}
  \end{equation} 
  where 
 \begin{eqnarray} 
 \varrho^2 &=& r^2+a^2p^2 \,, \nonumber\\
 \Omega &=& 1-\alpha pr \,, \nonumber\\ 
 Q&=& (1-\alpha^2r^2) (a^2+e^2+g^2-2mr+r^2) \,, \nonumber\\
 \tilde P &=& (1-p^2)\Big(1-2\alpha mp +\alpha^2(a^2+e^2+g^2)p^2\Big) \,, \nonumber\\ 
 F &=& (1+\alpha pr)\Big[\Big(1-\alpha^2(a^2+e^2+g^2)\Big)p
 +\alpha m(1-p^2)\Big] \nonumber\\ 
 &&\hskip3pc -2\alpha\Big[ r-(\alpha r-p)mp-\alpha(a^2+e^2+g^2)p^3\Big] \,, \nonumber\\ 
 G &=& (1+\alpha pr)\Big[\Big(1-\alpha^2(a^2+e^2+g^2)\Big)r
 -m(1-\alpha^2r^2)\Big] \nonumber\\ 
 &&\hskip3pc +2\alpha\Big[ (a^2+e^2+g^2)p+(\alpha r-p)mr-\alpha r^3\Big] \,. \nonumber
 \end{eqnarray} 
 Substituting for $r$ and $p$ ($=\cos\theta$) from (\ref{przerol}), these give the Weyl--Lewis--Papa\-petrou form of the solutions which represent accelerating and rotating black holes, at least for the region in which $Q>a^2P$. For the non-rotating case in which $a=0$, this transformation has been given by Hong and Teo \cite{HongTeo03} using an older coordinate system.

\section{Extending the Weyl coordinates up to the horizons}
\label{extension}

As already noted, the form of the metric (\ref{WeylMetric}) with (\ref{Weyltrans}) only describes that part of the stationary region for which $Q>a^2P$. For the rotating case for which $a\ne0$, it does not extend to the horizons on which $Q=0$. For the Kerr--Newman space-times, the metric thus applies only outside the ergosphere. This is bounded by a surface on which $e^{2\bar U}=0$, and which in the nonaccelerating case appears from infinity to be a surface of infinite redshift. However, the metrics (\ref{PleDemMetric}) and (\ref{newMetric}) already cover the space-time through these regions near the black hole. 
At this point, it is more significant that the metric functions (\ref{Weyltrans}) do not apply near the acceleration horizon at which also $Q=0$. In this section, we consider how to extend the metric up to the acceleration horizon.

To achieve this, we make use of the fact that there is an ambiguity in the $\bar t$ and $\bar\phi$ directions of the Killing vectors in the Weyl--Lewis--Papapetrou form of the metric. Following \cite{BicPra99} and \cite{Pravdas02}, we therefore introduce the coordinate transformation 
 $$ \bar t=\kappa_1\tilde t+\kappa_2\tilde\phi, \qquad 
 \bar\phi=\kappa_3\tilde t+\kappa_4\tilde\phi, $$ 
 in which the constant coefficients satisfy the condition
 $$ \kappa_1\kappa_4-\kappa_2\kappa_3=1. $$ 
 Applying this to the metric (\ref{WeylMetric}), leads to the line element 
 \begin{equation} 
 \d s^2= e^{2\tilde U}(\d\tilde t+\tilde A\,\d\tilde\phi)^2
 -e^{-2\tilde U}\Big[ e^{2\tilde\nu}
 (\d\brho^2+\d\bz^2)+\brho^2\d\tilde\phi^2 \Big] \ , 
 \label{Weyltilde} 
 \end{equation}  
 where 
 \begin{equation} 
 \begin{array}{rll}
 e^{2\tilde U} &=& (\kappa_1+\kappa_3\bar A)^2\,e^{2\bar U}
 -\kappa_3^2\,\brho^2\,e^{-2\bar U} \,,\nonumber \\[3pt] 
 \tilde A &=& {\displaystyle 
 {(\kappa_1+\kappa_3\bar A)(\kappa_2+\kappa_4\bar A)\,e^{2\bar U}
 -\kappa_3\kappa_4\,\brho^2\,e^{-2\bar U} 
 \over (\kappa_1+\kappa_3\bar A)^2\,e^{2\bar U}
 -\kappa_3^2\,\brho^2\,e^{-2\bar U}} } \,,\nonumber \\[12pt] 
 e^{2\tilde U}\,e^{-2\tilde\nu} &=& e^{2\bar U}\,e^{-2\bar\nu} 
 \,. \nonumber
 \end{array}
 \label{tildefns} 
 \end{equation}

It is now necessary to choose the parameters $\kappa_i$ to ensure that the function $e^{2\tilde U}$ does not become negative near the acceleration horizon on which $Q=0$. To achieve this, it is necessary that 
 \begin{equation} 
 \kappa_1+\kappa_3\,\bar A_{ah}=0, 
 \label{kappacond1} 
 \end{equation}  
 where $\bar A_{ah}$ is the value of $\bar A$ on the acceleration horizon where $\alpha r=\omega(|a|+l)^{-1}$ namely: 
 $$ \bar A_{ah}=-{1\over a} 
 \left({\omega^2\over\alpha^2(|a|+l)^2}+(a+l)^2\right). $$ 
 In fact, with this condition, 
 \begin{equation} 
 e^{2\tilde U}\approx {\kappa_3^2\,\varrho^2\over a^2\Omega^2}\>Q \,, 
 \label{horizfn1} 
 \end{equation}  
 evaluated near the acceleration horizon. This remains positive up to the horizon on which it vanishes. If fact, since this result has been obtained by considering the limit as $Q\to0$, the resulting metric applies to the entire stationary region between both horizons. i.e. with the condition (\ref{kappacond1}), the Weyl--Lewis--Papapetrou form of the metric is extended both out to the acceleration horizon and in through the ergosphere to the outer black hole event horizon.

In particular, for the physically most interesting case in which $l=0$, we have
$\bar A_{ah}=-(1+\alpha^2a^2)/\alpha^2a$, and hence we obtain 
  $$ \bar A-\bar A_{ah} ={(1+\alpha^2a^2p^2)Q-a^2(1-\alpha^2r^2)\tilde P
  \over \alpha^2a(Q-a^2\tilde P)} \,. $$ 
 Using this, and after cancelling a common factor $Q-a^2\tilde P$, we obtain 
 \begin{equation} 
 e^{2\tilde U}={\kappa_3^2\over \alpha^4 a^2}
{(1+\alpha^2a^2p^2)^2 Q - a^2(1-\alpha^2r^2)^2 \tilde P \over (r^2+a^2p^2)(1-\alpha pr)^2} \,, 
 \label{tildeU} 
 \end{equation} 
 and similarly 
 \begin{equation} 
 \tilde A =-\alpha^2a^2{\kappa_4\over\kappa_3}
 \left[{(1-p^2)(1+\alpha^2a^2p^2)Q-(a^2+r^2)(1-\alpha^2r^2)\tilde P
 \over (1+\alpha^2a^2p^2)^2Q-a^2(1-\alpha^2r^2)^2\tilde P }\right] \,, 
 \label{tildeA} 
 \end{equation} 
 in which an obvious factor of $Q$ may be cancelled. The remaining metric function $e^{2\tilde U}e^{-2\tilde\nu}$ is given, according to (\ref{tildefns}), by the last expression in (\ref{Weyltrans2}).

\section{Boost-rotation symmetric coordinates}

Following \cite{Bonnor83} and \cite{BicPra99}, we now make the further transformation to coordinates which extend across the acceleration horizon and in which the boost-rotation symmetry of the metric can be seen explicitly. Using these coordinates, the global properties as described in general terms in \cite{BicakSchmidt89} will become manifest. 

With the metric in the form (\ref{Weyltilde}), we now perform the transformation 
 \begin{equation} 
 \left.\begin{array}{l}
 \tilde t=\beta^{-1}\tanh^{-1}(t/z) \\[4pt]
 \brho=\gamma\,\rho\,\sqrt{z^2-t^2} \\[4pt]
 \bz=\bz_3-{1\over2}\gamma(\rho^2+t^2-z^2) 
 \end{array}  
 \right\} \quad \left\{  
 \begin{array}{l}
 \sqrt\gamma\,t=\pm\sqrt{\sqrt{\brho^2+(\bz-\bz_3)^2}
 +(\bz-\bz_3)}\>\sinh\beta\tilde t \\[4pt]
 \sqrt\gamma\,\rho= \ \sqrt{\sqrt{\brho^2+(\bz-\bz_3)^2}
 -(\bz-\bz_3)} \\[4pt]
 \sqrt\gamma\,z= \pm\sqrt{\sqrt{\brho^2+(\bz-\bz_3)^2}
 +(\bz-\bz_3)}\>\cosh\beta\tilde t
 \end{array} \right. 
 \label{b-r-trans} 
 \end{equation} 
 with $\tilde\phi=\phi$ and where $\beta$ and $\gamma$ are parameters whose values will be determined below. This takes the metric to the form 
 \begin{equation} 
 \begin{array}{l}
 {\displaystyle \d s^2= {e^\mu\over z^2-t^2}\bigg[ 
 (z\d t-t\d z) +A(z^2-t^2)\d\phi \bigg]^2 } \\[12pt]
 {\displaystyle \hskip7pc -e^\lambda\left[ {(z\d z-t\d t)^2\over z^2-t^2} +\d\rho^2 \right] 
 -e^{-\mu}\rho^2{\gamma^2\over\beta^2}\d\phi^2 , }
 \end{array}
 \label{br-metric} 
 \end{equation} 
 where 
 \begin{equation} 
 e^\mu={e^{2\tilde U}\over\beta^2(z^2-t^2)}, \qquad 
 A=\beta\,\tilde A, \qquad
 e^\lambda=\gamma^2\,{\rho^2+z^2-t^2\over e^{2\tilde U}e^{-2\tilde\nu}}. 
 \label{mulambda} 
 \end{equation}  
 To evaluate these functions explicitly, it is necessary to subsititute for $r$ and $\cos\theta=a^{-1}(\omega p-l)$ in expressions for $e^{2\tilde U}$, $\tilde A$ and $e^{2\tilde U}e^{-2\tilde\nu}$ in (\ref{tildefns}), (\ref{Weyltrans}) and (\ref{nu}) using (\ref{pr}). These will thus become dependent on the expressions for $R_i$ which, in terms of the boost-rotation coordinates, are directly given by 
 \begin{equation} 
 R_i=\sqrt{ \left({\textstyle{1\over2}}\gamma
 (\rho^2+z^2-t^2)-(\bz_i-\bz_3) \right)^2 +2\gamma\rho^2(\bz_i-\bz_3)} \,. 
 \label{Rinew} 
 \end{equation}  
 Note that, in particular, $R_3={1\over2}\gamma(\rho^2+z^2-t^2)$. Notice also that the boost is in the direction of the axis of symmetry.

Near the acceleration horizon where $Q\to0$, using (\ref{b-r-trans}), we find that 
 $$ z^2-t^2 \approx{\tilde P\,Q\over2\gamma\Omega^4|\bz-\bz_3|} \,, $$ 
 since $\bz-\bz_3<0$ in this case. It can thus be seen that the function $e^\mu$ given by (\ref{mulambda}) and (\ref{horizfn1}) is continuous across the acceleration horizon which is now located at $z^2-t^2=0$. It can similarly be seen that, near the horizon and with the condition (\ref{kappacond1}), $\tilde A\approx\kappa_4/\kappa_3$ which is a constant over the entire horizon. Finally, using (\ref{nu}) and (\ref{mulambda}), it is found that $e^\lambda$ is also continuous across this horizon.

It is now appropriate to choose the parameters $\kappa_i$, $\beta$ and $\gamma$ such that the metric (\ref{br-metric}) is asymptotically flat at spatial infinity where $\rho\to\infty$ for finite $z$ and $t$. This procedure will now be demonstrated explicitly for the physically significant case of accelerating Kerr--Newman solutions in which $l=0$.

Following \cite{BicPra99}, we consider the particular trajectory given in terms of Pleba\'nski--Demia\'nski coordinates by 
 $$ p=1-{1\over v^2}, \qquad \alpha r=1-{1\over v^4}, $$ 
 as $v\to\infty$ with $\bt$ and $\bphi$ fixed. In Weyl coordinates (\ref{transf1}), this trajectory behaves as 
 $$ \brho\approx2\,{b\over\alpha}\,v \,,\qquad 
 \bz\approx-{b\over\alpha}\,v^2 \,, $$ 
 where 
 $$ b=1-2\alpha m+\alpha^2(a^2+e^2+g^2) \,. $$ 
 In terms of the coordinates (\ref{b-r-trans}), in the limit, this trajectory behaves as
 $$ \rho\approx\sqrt{2\,b\over\alpha\gamma}\,v \,,\qquad 
 t\approx\pm\sqrt{2\,b\over\alpha\gamma}\,\sinh\beta\bt \,,\qquad 
 z\approx\sqrt{2\,b\over\alpha\gamma}\,\cosh\beta\bt \,, $$ 
 which indicates that it approaches spatial infinity as required. In order to determine the metric functions in this limit, we first calculate the asymptotic behaviour of various functions as follows. 
 $$ {Q-a^2\tilde P\over\varrho^2\,\Omega^2} \approx-{2\alpha^2a^2b\over1+\alpha^2a^2}\,v^2, \qquad
 \bar A-\bar A_{ah} \approx -{1+\alpha^2a^2\over\alpha^4a^3}\,{1\over v^2}, 
 \qquad e^{2\bar U}e^{-2\bar\nu}={2b^3\over1+\alpha^2a^2}\,v^2. $$ 
 Then, using (\ref{tildefns}) and (\ref{mulambda}), we find that the metric functions approach the following constants  
 \begin{equation}
 e^\mu \approx {\kappa_3^2\gamma(1+\alpha^2a^2)\over\beta^2\alpha^3a^2} \,,
 \qquad A \approx {\kappa_2\beta\alpha^2a\over\kappa_3(1+\alpha^2a^2)} \,, 
 \qquad e^\lambda \approx {\gamma(1+\alpha^2a^2)\over\alpha b^2} \,.
 \label{fnsatinfty}
 \end{equation}
 In order for the space-time to be asymptotically Minkowskian at spatial infinity, we require that $e^\mu\to1$, $A\to0$ and $e^\lambda\to1$. Taking these in reverse order, we first obtain that 
 \begin{equation} 
 \gamma={\alpha\,b^2\over1+\alpha^2a^2}, 
 \label{gamma} 
 \end{equation} 
  and then 
 \begin{equation} 
 \kappa_2=0 \,, 
 \label{kappa2} 
 \end{equation}  
 (which shows that the initial choice of $\bt$ in (\ref{trans1A}) was particularly convenient)
 and finally 
 \begin{equation} 
 \kappa_3={\beta\alpha a\over b} \,, 
 \label{kappa3} 
 \end{equation}  
 which with (\ref{kappacond1}) implies that 
 \begin{equation} 
 \kappa_1={1\over\kappa_4} ={\beta(1+\alpha^2a^2)\over\alpha b} \,.
 \label{kappa14} 
 \end{equation}  
 Thus, the condition of asymptotic flatness at spatial infinity fixes all the free parameters except for $\beta$.

The final parameter $\beta$ is determined by examining the regularity of the axis. In fact, the axis is regular if 
 $$ \beta^2=\gamma^2e^{-(\mu+\lambda)}|_{\rho=0} \,. $$ 
 A straightforward calculation shows that this regularity condition is explicitly 
 \begin{equation} 
 \beta=\gamma{1-2\alpha mp_0+\alpha^2(a^2+e^2+g^2) \over 
 1-2\alpha m\>+\alpha^2(a^2+e^2+g^2)} \,,
 \label{beta} 
 \end{equation}  
 where $p_0=1$ for regularity on the half-axis $\theta=0$, or $p_0=-1$ for regularity on $\theta=\pi$. It is thus clear that the axis cannot be regular everywhere when $\alpha\ne0$. We can therefore chose to satisfy (\ref{beta}) with $p_0=-1$ so that the axis between the two black holes is regular and the black holes are accelerated by conical singularities with a deficit angle given by  
 $$ \delta_0={8\pi\alpha m\over1+2\alpha m+a^2(a^2+e^2+g^2)} \,. $$ 
 This represents strings which pull the black holes towards infinity as described in \cite{GriPod05}.

The final form of the metric in boost-rotation-symmetric form is now given by the line element (\ref{br-metric}) in which the metric functions (\ref{mulambda}) are fully determined. For the case in which $l=0$, they can be obtained explicitly by substituting (\ref{tildeU}), (\ref{tildeA}) and the last expressions in (\ref{tildefns}) and (\ref{Weyltrans2}) directly into (\ref{mulambda}), where $\beta$, $\gamma$ and $\kappa_i$ are given by (\ref{beta}), (\ref{gamma}), (\ref{kappa14}), (\ref{kappa2}) and (\ref{kappa3}). It is then necessary to substitute for $p$ and $r$ from (\ref{przerol}). The resulting expressions are extremely cumbersome. However, by using the identity (\ref{Risquared}), it is found that 
 \begin{equation} 
 1-\alpha pr ={4\bz_1(\bz_3^2-\bz_1^2) \over 
 m(\bz_1+\bz_3)R_1 +m(\bz_1-\bz_3)R_2
 +2\bz_1({\textstyle{1\over\alpha}}+\bz_3)R_3 -2\bz_1(\bz_1^2-\bz_3^2)} \,, 
 \label{Omega} 
 \end{equation} 
 in which the denominator is identical to that of the expressions for $p$ and $\alpha r$ in (\ref{przerol}). In addition, we can substitute 
 $\rho^2+z^2-t^2=2R_3/\gamma$ and 
 \begin{equation} 
 z^2-t^2={(R_1+R_3+\bz_1-\bz_3)(-R_1+R_3+\bz_1-\bz_3) \over 
2\gamma(\bz_1-\bz_3)} \,. 
 \label{ztsquared} 
 \end{equation}  
  Again, the identity (\ref{Risquared}) permits considerable simplifications. Nevertheless, the resulting expressions remain very lengthy. We will therefore state the result explicitly only for the special case of the $C$-metric for which remarkable simplifications can be achieved.

Before giving this result, however, it may be noticed that this procedure is effectively a coordinate transformation directly from Pleba\~nski--Demia\~nski coordinates to those of the boost-rotation-symmetric form. i.e. it explicitly avoids using the Weyl--Lewis--Papapetrou coordinates which apply only in the stationary region.

\subsection{The $C$-metric} 

To illustrate the above procedure, we will explicitly derive the boost-rotation-symmetric form of the $C$-metric for which the parameters $e$, $g$, $a$, and $l$ all vanish. i.e. we consider an accelerating black hole of mass $m$.

For this case, using (\ref{tildeU}), (\ref{tildeA}) and (\ref{Weyltrans2}) with (\ref{przerol}) and (\ref{Omega}) according to the above method, we obtain 
 $$ \begin{array}{l}
 e^\mu ={\displaystyle {\Big[-R_1+R_3+{1\over2\alpha}+m\Big] 
 \Big[R_2+R_3+{1\over2\alpha}-m\Big] \Big[R_1+R_2-2m\Big] \over
 b^2\ (z^2-t^2)\ \Big[(1-2\alpha m)R_1+(1+2\alpha m)R_2+4\alpha mR_3\Big]} 
 \,, }\\[18pt]
 e^\lambda={\displaystyle {\alpha b^4(\rho^2+z^2-t^2)
 \Big[(1-2\alpha m)R_1+(1+2\alpha m)R_2+4\alpha mR_3\Big]^2 \over
 8\,(1-4\alpha^2m^2)^2\,R_1\,R_2\,R_3} }\,,
 \end{array} $$ 
 and $A=0$. Using (\ref{Rinew}), these express the metric functions explicitly in terms of the correct coordinates. However, using the identity (\ref{ztsquared}) together with (\ref{Risquared}), these expressions reduce to the remarkably simple forms 
 \begin{equation} 
 e^\mu ={ R_1+R_2-2m \over R_1+R_2+2m} \,,\qquad 
 e^\lambda= { \Big[(1-2\alpha m)R_1+(1+2\alpha m)R_2+4\alpha mR_3\Big]^2 \over
 4\,(1+2\alpha m)^2\,R_1\,R_2}  \,. 
 \end{equation}

It may be noted that these expressions differ from those presented elsewhere \cite{Bonnor83, Pravdas00}. This arises since the roots $\bz_i$ and the associated cubic are different. Here, we have simpler roots and simpler final expressions, with the added feature that an extension to the rotating case is also obtained.

\section{Conclusion}

The metric (\ref{newMetric}) which was presented in \cite{GriPod05} nicely describes most of the properties of an accelerating Kerr--Newman--NUT black hole. In particular, it clearly represents the interior structure of the black hole, its horizon structure out to and through the acceleration horizon, and the conical singularity that is needed to produce the acceleration. However, it does not describe the global structure of the complete space-time beyond the acceleration horizon. In particular, it does not show that the analytically extended space-time actually contains two causally separated black holes which accelerate away from each other.

To demonstrate this particular property, we first cast the metric in the Weyl--Lewis--Papapetrou form. However, it was observed that the transformation (\ref{transf1}) of the metric (\ref{newMetric}) to the Weyl--Lewis--Papapetrou form is only valid in a restricted part of the stationary region. In section~\ref{extension}, we have identified the unique transformation to this form of the metric which covers the entire stationary region between the outer black hole horizon and the acceleration horizon. Then, using the transformation (\ref{b-r-trans}), we have further transformed the metric to a boost-rotation-symmetric form (\ref{br-metric}). In this form, the acceleration horizon is represented by the null hypersurfaces on which $z^2-t^2=0$, and the metric can be extended through these horizons to the full infinite $t$-$z$ plane. 
The space-time represented by this metric contains two distinct stationary regions, in which $z$ has different signs. These regions are mirror images of each other and each contains a black hole-like source. Thus, when $\alpha\ne0$ the complete space-time represented by (\ref{br-metric}) actually contains two causally separated accelerating black holes as illustrated in figure~1. This is consistent with the general global properties of boost-rotation symmetric space-times as described in~\cite{BicakSchmidt89}.

\begin{figure}[hbt]
\begin{center} \includegraphics[scale=0.55, trim=5 5 5 -5]{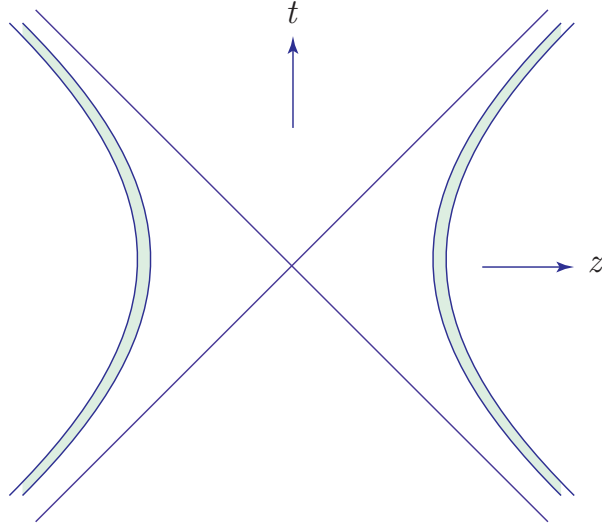} 
\caption{ \small A schematic diagram representing the space-time outside a pair of accelerated black holes. The acceleration horizons occur at $t=\pm z$, while the shaded regions represent the outer black hole horizons. Obviously, these horizons should be null but, in the weak field limit, the regions indicated represent the trajectories of the source particles.  }
\label{structure}
\end{center}
\end{figure}

\section*{Acknowledgements}

This work was partly supported by a grant from the EPSRC.

\end{document}